\documentclass{article}%
\usepackage{amsmath}
\usepackage{amsfonts}
\usepackage{amssymb}
\usepackage{graphicx}
\usepackage{float}
\usepackage{subcaption}
\usepackage{multirow}
\usepackage{color}
\usepackage{longtable}
\usepackage[dvipsnames]{xcolor}
\usepackage{amsmath,amssymb,amsthm}
\usepackage[margin=1.0in]{geometry}
\usepackage{cite}
\usepackage{graphicx}
\usepackage{float}
\usepackage{subcaption}
\usepackage{enumerate}
\usepackage{subcaption}
\usepackage{stackengine}
\allowdisplaybreaks
\usepackage[colorlinks=true
,urlcolor=blue
,anchorcolor=blue
,citecolor=blue
,filecolor=blue
,linkcolor=blue
,menucolor=blue
,pagecolor=blue
,linktocpage=true
,pdfproducer=medialab
,pdfa=true
]{hyperref}
\newcommand{\bpartial}{\mathop{\partial\kern -4pt\raisebox{.8pt}{$|$}}}
\newcommand{\bra}{\mathopen{[\kern-1.6pt[}}
\newcommand{\ket}{\mathclose{]\kern-1.5pt]}}
\newcommand{\bbra}{\mathopen{[\kern-2.2pt[\kern-2.3pt[}}
\newcommand{\bket}{\mathclose{]\kern-2.1pt]\kern-2.3pt]}}

\makeindex
\begin{document}
	\title {\large{ \bf 
			Regular  $(2+1)$-dimensional spatially homogeneous $\alpha'$-corrected BTZ-like black hole in string theory}}
	
	\vspace{3mm}
		\author {  \small{ \bf  F. Naderi}\hspace{-1mm}{ \footnote{%
					 e-mail:
			 f.naderi@azaruniv.ac.ir}} ,{ \small	} \small{ \bf  A. Rezaei-Aghdam}\hspace{-1mm}{
		\footnote{Corresponding author,
			 e-mail:	rezaei-a@azaruniv.ac.ir}} \\
		 		{\small{\em
			Department of Physics, Faculty of Basic Sciences, Azarbaijan Shahid Madani University}}\\
	{\small{\em   53714-161, Tabriz, Iran  }}}

\maketitle

\begin{abstract}

   We consider a  $(2+1)$-dimensional spacetime whose  two-dimensional space part is Weyl-related to a surface of arbitrary negative constant  Gaussian curvature with symmetries of two-dimensional Lie algebra. It is shown that the geometry is a Lobachevsky-type geometry described by deformed hyperbolic function. At leading order string effective action with the source given by dilaton and antisymmetric  $B$-field in the presence of central charge deficit term $\Lambda$, we obtained a solution whose line element is Weyl-related to this homogeneous spacetime with arbitrary negative Gaussian curvature. The solution can be transformed to the BTZ-like black hole by coordinate redefinition, while  the BTZ black hole can be recovered by choosing a special set of parameters. The solutions appear to be in the high curvature limit $R\alpha'\gtrsim1$, with emphasis on including the higher order $\alpha'$ corrections. Considering the two-loop (first order $\alpha'$) $\beta$-function equations of $\sigma$-model, we also present the $\alpha'$-corrected black hole solutions.

\end{abstract}
\section{Introduction}
\subsection{General considerations}
Investigating the $(2+1)$-dimensional gravity theories, initiated by \cite{staruszkiewicz1963gravitation}, have attracted attention especially as
a toy model of quantum gravity, motivated by the background provided in \cite{deser1984three,deser1988classical,t1988non}, to
survey the classical and quantum dynamics of point sources, in addition to representation of Chern-Simons theory for $(2+1)$-dimensional gravity   \cite{WITTEN1989113,WITTEN19,witten1991quantization}.  
Classical and quantum solutions for gravity theories in this dimension have been widely investigated, for instance, in \cite{21Hamber2012,21Carlip1994,Darabi2013,PhysRevD.90.084008,Naderi_2021,Adami2020,EGHBALI2017791,Darabi_2022}.

 The first exact black hole solution for $(2+1)$-dimensional gravity with a negative cosmological constant,  called BTZ black hole solution, was first obtained in  \cite{PhysRevLett.69.1849,PhysRevD.48.1506}. A modified BTZ black hole solution was then obtained to a $(2 + 1)$-dimensional string theory with a matter source given by anti-symmetric $B$-field with the contribution of the central charge deficit $\Lambda$ \cite{Horowitz}. BTZ black hole solutions have been widely improved,  generalized, and investigated from different physical viewpoints,  to extend the quantum theory of gravity, gauge field theory, and string theory, along with improving the knowledge of gravitational interaction in lower dimensional manifolds \cite{Darabi2013,EGHBALI2017791,Carlip2005,SCarlip,cataldo2001btz,044027,sheykhi2021mimetic}. Also, the role of BTZ black hole in making deformed graphene, which is a particular $2$-dimensional real system \cite{Semenoff:1984dq}, as a tabletop for QFT in curved spacetimes to study the measurable effects,  like Hawking-Unruh effect has been pointed out in \cite{g}. 
 An important role, in this case, was played by surfaces of constant negative Gaussian curvature, whose embedding into ${ R}^3$ gives rise to essential singularities, as a result of the Hilbert theorem  \cite{eisenhart,14}. 
 In this category, the line element of $(2+1)$-dimensional spacetimes constructed by Beltrami, hyperbolic, and elliptic pseudospheres have been shown to be Weyl-related to the line element of Rindler, de Sitter, and non-rotating BTZ black holes. 

In this work,  we extend the above catalog with a  $(2+1)$-dimensional homogeneous spacetime, whose spatial part is Weyl-related to a surface of negative constant Gaussian curvature with the symmetries of two-dimensional Lie algebra, admitting a homogeneous metric.
The homogeneous spacetimes,    which possess the symmetry of spatial homogeneity and are defined based on the simply-transitive  Lie groups classification \cite{Cosmictopology},  have been  used to construct cosmological and black hole solutions in $(4+1)$  dimensions \cite{bh,MOJAVERI012}, $(3+1)$  dimensions \cite{Ellis1969,batakis2,PhysRevD.57.5108,NADERI2017,Naderi2}, and $(2+1)$ dimensions \cite{Naderi_2021}.
Here, we show that the considered  homogeneous surface, being an especial case of Lobachevsky geometries, can be reduced to a hyperbolic surface if the Gaussian curvature is set to $-\frac{1}{2}$. 
In a general case with an arbitrary negative Gaussian curvature, considering a $(2+1)$-dimensional spacetime, 
which is conformal related to this homogeneous spacetime,  we  find  solutions for the leading order of string effective action including dilaton, field strength tensor of $B$-field, and the central charge deficit term $\Lambda$  that appears as a negative cosmological constant. 

It turned out that the string frame metric is in the high curvature limit, i.e.  $R\alpha'\gtrsim 1$,  where the $\alpha'$ is square of string length,  $\alpha'=\lambda_s^2/2\pi$. At this limit, the role of higher-order $\alpha'$ corrections to the string effective action becomes significant, which is widely believed to regularize the curvature singularity  \cite{brustein1994graceful}. Although considering all orders in $\alpha'$ is recommended at the high curvature regime, we limit the calculations to the first order of $\alpha'$, hoping to present a glimpse of the consequences that could be obtained considering all $\alpha'$ orders.
Considering only the first order of corrections, 
the regularizing effects of the $\alpha'$ corrections have been already investigated in \cite{gasperini1997towards,cartier2001gravitational,niz2007stringy,BF2,Eghbali_2021}, while the higher orders of $\alpha'$ corrections usually result in the order by order correcting the lower order solutions.
Similar to what we have done in \cite{NADERI2017,Naderi2}, implementing a perturbative series expansion in $\alpha'$ on the background field for string effective action that includes the correction of Gauss-Bonnet type, we find the $\alpha'$ correction to the obtained black hole solutions.

In the following, we add some introductory remarks on  two-loop $\beta$-functions of $\sigma$-model, which are equivalent to the $\alpha'$-corrected string effective action equations of motion, followed by a brief review on geometries with constant Gaussian curvature.

\subsection{String effective action and two-loop $\beta$-functions }\label{tsec2}

We are going to construct black hole solutions to string effective action, whose equations of motion are equivalent to the $\beta$-function equations of $\sigma$-model, which assuring the conformal invariance of the $\sigma$-model, are, on the other hand,  equivalent to the field equations of the associated gravity theory  \cite{tseytlin1992elements}.  
For a $\sigma$-model  with  background fields of metric $g$, dilation field $\phi$, and antisymmetric $B$-field,  the two-loop $\beta$-functions (order  $\alpha'$) have been calculated in \cite{metsaev1987order,HULL1988197}, where the $\beta$-function equations for the metric are
\begin{equation}\label{tbetaGR}
\begin{aligned}
\frac{1}{\alpha'}{\beta}^{g}_{\mu\nu}=&{R}_{{\mu}{\nu}}-\frac{1}{4}{H}^{2}_{{\mu}{\nu}}-\nabla_{{\mu}}\nabla_{{\nu}}{\phi}+\frac{\alpha'}{2}\big[
R_{\mu\alpha\beta\gamma}R_{\nu}^{~\alpha\beta\gamma}
-\frac{3}{2}R_{(\mu}^{~~~\alpha\beta\gamma}H_{\nu)\alpha\lambda}H_{\beta\gamma}^{~~\lambda}
-\frac{1}{2}R^{\alpha\beta\rho\sigma}H_{\mu\alpha\beta}H_{\nu\rho\sigma}        +\frac{1}{8}(H^4)_{\mu\nu}
\\
&-\frac{f}{2}\big(R_{\mu\alpha\beta\nu}(H^2)^{\alpha\beta}
+2\,R_{(\mu}^{~~\alpha\beta\gamma}H_{\nu)\alpha\lambda}H_{\beta\gamma}^{~~\lambda}
+R^{\alpha\beta\rho\sigma}H_{\mu\alpha\beta}H^{\nu\rho\sigma}
-\nabla_{\lambda}H_{\mu\alpha\beta}\nabla^{\lambda}H_{\nu}^{~\alpha\beta}
\big)    \\
&-\frac{1}{12}\nabla_{\mu}\nabla_{\nu}H^2
+\frac{1}{4}\nabla_{\lambda}H_{\mu\alpha\beta}\nabla^{\lambda}H_{\nu}^{~\alpha\beta}
+\frac{1}{12}\nabla_{\mu}H_{\alpha\beta\gamma}\nabla_{\nu}H^{\alpha\beta\gamma}
+\frac{1}{8}H_{\mu\alpha\lambda}H_{\nu\beta}^{~~\lambda}(H^2)^{\alpha\beta}\big],
\end{aligned}
\end{equation}
where  $H^4=H_{\mu\nu\lambda}H^{\nu\rho\kappa}H_{\rho\sigma}^{~~\lambda}H^{\sigma\mu}_{~~\kappa}$,  $H_{\mu\nu}^2=H_{\mu\rho\sigma}H_{\nu}^{\rho\sigma}$   and $H$ is  the field strength of $B$-field defined by $H_{\mu\nu\rho}=3\partial_{[\mu}B_{\nu\rho]}$. The $f$ parameter stands for the renormalization scheme (RS) dependence in $\beta$-functions and the corresponding schemes to  $f=1$ and $f=-1$, called $R^2$ and {Gauss-Bonnet} schemes,    have been particularly discussed in \cite{metsaev1987order}. Solutions of various RS $\beta$-function equations are different, but still equivalent because they belong to various definitions of physical metric, dilaton, and $B$-field. We are going to focus on the Gauss-Bonnet scheme, where the $\beta$-function of $B$-field is given by \cite{metsaev1987order}
\begin{equation}\label{tbetaBG}
\begin{aligned}
\frac{1}{\alpha'}{\beta}^{B}_{\mu\nu}(f&=-1)=\hat{R}_{[{\mu}{\nu}]}+\frac{\alpha'}{4}\big(2\hat{R}^{\alpha\beta\gamma}_{~~~~[\nu}\hat{R}_{\mu]\alpha\beta\gamma}-\hat{R}^{\beta\gamma\alpha}_{~~~~[\nu}\hat{R}_{\mu]\alpha\beta\gamma}+\hat{R}_{\alpha[\mu\nu]\beta}H^{\alpha\beta}-\frac{1}{12}H_{\mu\nu}^{~~\rho}\nabla_{\rho}H^2\big),
\end{aligned}
\end{equation}
in which   the $\hat{R}^{\rho}_{\mu\nu\sigma}$ is the Riemann tensor of the generalized connection with torsion ${\hat{\Gamma}}^{\rho}_{\mu\nu}={{\Gamma}}^{\rho}_{\mu\nu}-\frac{1}{2}H^{\rho}_{\mu\nu}$.\footnote{
	${\hat{R}}^{\kappa}_{~\lambda\mu\nu}= R^{\kappa}_{~\lambda\mu\nu}-\frac{1}{2}\nabla_{\mu}H^{\kappa}_{\nu\lambda}-\frac{1}{2}\nabla_{\nu}H^{\kappa}_{\mu\lambda}+\frac{1}{4}H_{~\nu\lambda}^{\gamma}H_{~\mu\gamma}^{\kappa}-\frac{1}{4}H_{~\mu\lambda}^{\gamma}H^{\kappa}_{~\nu\gamma}.
	$} For dilation field,
the averaged    $\beta$-function, which is given in terms of metric and dilation  $\beta$-functions as $   \tilde{\beta}^{{\phi}}={\beta}^{{\phi}}-\frac{1}{4}{\beta}^{g}_{\mu\nu}g^{\mu\nu}$, is given  by \cite{metsaev1987order}
\begin{eqnarray}\label{tbetafR}
\begin{split}
\frac{1}{\alpha'}\tilde{\beta}^{{\phi}}&=-{R}+\frac{1}{12}{H}^{2}+2\nabla_{{\mu}}\nabla^{{\mu}}{\phi}+(\partial_{{\mu}}{\phi})^{2}-\Lambda-\frac{\alpha'}{4}(R^2_{\mu\nu\rho\lambda}-\frac{1}{2}R^{\alpha\beta\rho\sigma}H_{\alpha\beta\lambda}H_{\rho\sigma}^{~~\lambda}
+\frac{1}{24}H^4-\frac{1}{8}(H_{\mu\nu}^{~~2})^2),
\end{split}
\end{eqnarray} 
in which $\Lambda$ indicates the central charge deficit of  $D$-dimensional bosonic theory, given by   \cite{tseytlin1992elements} 
\begin{eqnarray}\label{lamda}	\Lambda=\frac{2\,(26-D)}{3\alpha'}.
\end{eqnarray}

The $\beta$-function equation \eqref{tbetafR} can be obtained   
by variation of the following string effective action with respect to the dilaton field \cite{metsaev1987order}
\begin{eqnarray}\label{taction}
\begin{split}
S=-\frac{1}{2\kappa_D^2}\int &d^D x\sqrt{g}e^{\phi}(R-\frac{1}{12}H^2+(\nabla\phi)^2+\Lambda+\frac{\alpha'}{4}(R^2_{\mu\nu\rho\lambda}-\frac{1}{2}R^{\alpha\beta\rho\sigma}H_{\alpha\beta\lambda}H_{\rho\sigma}^{~~\lambda}
+\frac{1}{24}H^4-\frac{1}{8}(H_{\mu\nu}^{~~2})^2)),
\end{split}
\end{eqnarray}  
in which  
$\kappa_D^2=8\pi G_D=\lambda_s^{D-2}{\rm e}^{-\phi}=\lambda_p^{D-2}$, where $\lambda_p$ is the Planck length and 
$G_D$ is the $D$-dimensional gravitational Newton constant.

The string frame metric $g_{\mu\nu}$ is related to  the   Einstein-frame metric  $\tilde{g}_{\mu\nu}$  in $D$-dimensional spacetime by
\begin{equation}\label{tcon}
\tilde{	g}_{\mu\nu}={\rm e}^{\frac{2}{D-2}\phi}g_{\mu\nu}.
\end{equation}
To provide the Einstein-frame string effective action, a remained intrinsic ambiguity to a given order $\alpha'$  in the effective action was  considered in \cite{metsaev1987order}. This ambiguity is the result of the invariance of the  $S$-matrix under a set of field redefinitions \cite{Metsaev1987}, which leads to a class of physically equivalent effective actions parametrized by $8$   essential coefficients  \cite{Tseytlin2}. The Gauss-Bonnet scheme has been used in \cite{metsaev1987order}
to provide a set of these  $8$ coefficients.  In $3$ dimensions  the  effective action is given by 
\begin{eqnarray}\label{tGBaction}
\begin{aligned}
S=&-\frac{1}{2\kappa_3^2}\int d^3 x\sqrt{\tilde{	g}}\bigg(\tilde{R}-\frac{1}{12}\,{\rm e}^{{4}{}\phi}H^2-(\tilde{\nabla}\phi)^2+\Lambda {\rm e}^{-2\phi}+\frac{\alpha'e^{{2}{}\phi}}{4}\bigg[\tilde{R}^2_{\mu\nu\rho\lambda}-4\,\tilde{R}_{\mu\nu}^2+\tilde{R}^2-\left(\partial\phi\right)^4\\
&+{\rm e}^{{4}\phi}\left(-\frac{1}{2}\tilde{R}^{\alpha\beta\rho\sigma}H_{\alpha\beta\lambda}H_{\rho\sigma}^{~~\lambda}+H^2_{\mu\nu}\tilde{\nabla}^{\mu}\phi\tilde{\nabla}^{\nu}\phi
-\frac{1}{6}H^2(\tilde{\nabla}\phi)^2\right)+{\rm e}^{{8}{}\phi}\left(\frac{1}{24}H^4+\frac{1}{8}(H_{\mu\nu}^{~~2})^2-\frac{1}{16}(H^2)^2\right)\bigg]\bigg),
\end{aligned}
\end{eqnarray}
in which  $\tilde{\nabla}$ indicates the covariant derivative with respect to  $\tilde{g}_{\mu\nu}$. 

The $\alpha'$ corrections to the string effective action \eqref{taction} are significant when the curvature is at the high limit $R\alpha'\gtrsim 1$. 
Actually, the string effective action is known to include two kinds of corrections:  the stingy type $\alpha'$-expansion  at high curvature regime and the quantum nature loop expansion in string coupling 
at strong string coupling limit $g_s\equiv {\rm{e}^{-\phi}}>1$ \cite{gasperini1997towards}. As long as the $g_s$ is sufficiently weak in the high curvature regime, the loop corrections are allowed to be neglected, and the $\alpha'$ corrections are enough to be included in the effective action  \cite{gasperini1997towards}.

\subsection{Lobachevsky geometry}
The surfaces of constant Gaussian curvature have attracted attention in the classic studies of differential geometry \cite{eisenhart,spivak}. Particularly, these spacetimes have been considered as real terrestrial laboratories in studying the measurable effects, like the Hawking-Unruh effect \cite{g}. As a consequence of a  theorem proven by Hilbert \cite{eisenhart,14,15}, embedding of these surfaces into $R^3$ gives rise to essential singularities. Although no single good parametrization exists for all surfaces, the surfaces of revolution can be described by one such parametrization,  namely the canonical parametrization.\footnote{The surfaces of revolution are the swapped surfaces by a  curve for instance in the plane $(x,z)$  rotated around the $z$-axis by a full angle.} These surfaces can be parameterized in ${ R}^3$ by \cite{g}
\begin{equation} \label{canonicalpar}
x(\rho,v) = R(\rho) \cos v \;, \; y(\rho,v) = R(\rho) \sin v \; , \; z(\rho) = \pm \int^\rho \sqrt{1 - {R'}^2(\bar{\rho})} d \bar{\rho} \;,
\end{equation}
where $R(\rho)$ identifies the type of surface,  $\rho$ is the meridian coordinate, whose range is fixed by request of real $z (\rho)$, $v$ is the parallel coordinate (angle) ranging in $ [0, 2 \pi]$, and
prime denotes derivative with respect to $\bar{\rho}
$.
 The parametrization \eqref{canonicalpar} leads to the following
 embedded line element 
\begin{equation}\label{linelsurfrev}
d l^2 \equiv dx^2 + dy^2 + dz^2 = d\rho^2 + {R}^2(\rho) dv^2.
\end{equation}
The Gaussian curvature of this line element is ${\cal K} = - \frac{R''(\rho)}{R(\rho)}$  \cite{spivak},
which can be solved as a differential equation for positive and negative constant ${\cal K}$, giving the following  cases
\begin{equation}\label{solgaussian1}
R(\rho) = c \cos (\sqrt{{k}}\,\rho  ) \quad {\rm for} \quad {\cal K} = k \;,
\end{equation}
\begin{equation}\label{solgaussian2}
R(\rho) = c_1 \sinh (\sqrt{k}\,\rho  ) + c_2 \cosh (\sqrt{k}\,\rho )\quad {\rm for} \quad {\cal K} = - k \;,
\end{equation}
where $k$ (positive), $   c, c_1$, and $ c_2$ are  real constants.

In the positive constant curvature case  \eqref{solgaussian1},  three   cases  are usually distinguished by $c = k^{-\frac{1}{2}}$, $c > k^{-\frac{1}{2}}$, $c < k^{-\frac{1}{2}}$, which are describing a sphere of radius $k^{-\frac{1}{2}}$, and two applicable surfaces   to  sphere via a redefinition of $v \to (c\sqrt{k}) v$. On the other hand, for negative constant curvature class   \eqref{solgaussian2}, called the  Lobachevsky surfaces \cite{g}, the  three special cases of Beltrami, elliptic, and hyperbolic pseudospheres have attracted attention, which are described as follows:

$\bullet$ The  Beltrami pseudosphere, whose associated spacetime is known to be conformal to the Rindler spacetime \cite{g}, 
is defined by \eqref{solgaussian2} with $c_1 = c_2 \equiv c$, and
\begin{equation} \label{Rbeltrami}
R(\rho) = c \; e^{\sqrt{k}\,\rho},  \quad {\rm where} \quad R(\rho) \in [0, k^{-\frac{1}{2}}] \Leftrightarrow \rho \in  [- \infty,- k^{-\frac{1}{2}}\ln(c\sqrt{k})],
\end{equation}
in which $c$ is required to be a  real positive number.

$\bullet$ With  $c_2 = 0$, $c_1 \equiv c$, the equation \eqref{solgaussian2} describes the elliptic pseudosphere  by
\begin{equation} \label{Relliptic}
R(\rho) = c \; \sinh (\sqrt{k}\,\rho), \quad {\rm where} \quad R(\rho) \in [0, k^{-\frac{1}{2}}\cos \vartheta] \Leftrightarrow \rho  \in  [0 , {\rm arcsinh} \cot \vartheta],
\end{equation}
in which $\sin(\vartheta)=c\sqrt{k}$.
The spacetime whose space part is an Elliptic surface has been shown to be Weyl to  de-Sitter spacetime \cite{iorio2015curved}.

$\bullet$ The  $c_1 = 0$, $c_2 \equiv c$ case leads to the hyperbolic pseudosphere with 
\begin{equation} \label{Rhyperbolic}
R(\rho) = c \; \cosh(\sqrt{k}\,\rho), \quad {\rm where} \quad R(\rho) \in [c, \sqrt{k^{-1} + c^2}] \Leftrightarrow \rho  \in  [- {\rm arccosh} (\sqrt{1 + \frac{1}{k\,c^2}}) , {\rm arccosh} (\sqrt{1 + \frac{1}{k\,c^2}})],
\end{equation}
in which $c$ is required to be a  real positive number.  
The interesting characteristic of hyperbolic pseudosphere is that  the line element of the non-rotating BTZ black hole  is Weyl-related to the line element of the hyperbolic  spacetime \cite{PhysRevLett.69.1849}, in such a way that \cite{cvetivc2012graphene}
\begin{equation} \label{BTZ}
\begin{split}
ds^2_{BTZ}&=-\left(\frac{r^2}{l^2}-m\right)\,dt^2+\frac{dr^2}{\left(\frac{r^2}{l^2}-m\right)}+r^2d\phi^2\\
&=\frac{l^2m}{\sinh^2\left(\sqrt{m}\rho\right)}\left(-dt^2+d\rho^2+l^2\cosh^2\left(\sqrt{m}\rho\right)d\phi^2\right),
\end{split}
\end{equation}
where the $r$  is given in terms of $\rho$ by $r=\sqrt{m}l\coth(\sqrt{m}\rho)$.


To find a black hole solutions for string effective action with a line element Weyl-related to a homogeneous spacetime with Lobachevsky geometry, the paper is organized as follows: 
Section \ref{se2} presents the characteristics of the considered spacetime in detail and then the derivation of the black hole solutions.
After introducing the geometry of the $(2+1)$-dimensional spacetime, the black hole solutions obtained by solving the one-loop order of $\beta$-functions, i.e. the zeroth order $\alpha'$, are provided in subsection \ref{tsec3}. The thermodynamic characteristics of these solutions are investigated in section \ref{thermo}. Then, section \ref{22} presents the $\alpha'$-corrected black hole solutions constructed based on the solutions of section \ref{se2}, by solving the two-loop order $\beta$-function equations in the Gauss-Bonnet RS. 
Finally, some concluding remarks are presented in section \ref{conclussion}. 

\section{Black hole solution of string effective action on spatially homogeneous $(2+1)$-dimensional spacetime}\label{se2}

 On the $(2+1)$-dimensional spacetime, whose  the $t$-constant hypersurface is given by a homogeneous space corresponding to the $2$-dimensional Lie  group with real two-dimensional Lie algebra $[T_1,T_2]=T_2$, the following metric ansatz  can be considered \cite{Naderi_2021}
\begin{eqnarray}\label{metricc}
ds^2=-dt^{2}+g_{ij}\sigma^{i}\sigma^{j},
\end{eqnarray}
where  $g_{ij}$ are constants and $\{\sigma^{i},~i=1,2\}$  are left invariant basis $1$-forms  {on the Lie group}, obeying $\sigma^{2}=-\frac{1}{2} 
\sigma^{1}\wedge \sigma^{2}$, $\sigma=(g^{-1}dg)^iT_i$, and $\sigma^i=(g^{-1}dg)^i$, where $g={\rm e}^{x_1T_1} {\rm e}^{x_2T_2}$. The  coordinate and non-coordinate basis are related by
\begin{eqnarray}\label{sigma}
\sigma^1=dx_{1}+
x_2dx_2,\quad \sigma^2=dx_2.
\end{eqnarray}
Accordingly, the  metric \eqref{metricc} recasts the following form
\begin{eqnarray}\label{m1}
ds^2=-dt^2+(g_{11}+2 g_{12}\,x_2+g_{22} \,x_2^2)\,dx_1^2+2\left(g_{12}+g_{22}x_2\right)\,dx_1dx_2+g_{22}\,dx_2^2,
\end{eqnarray}
whose Gaussian curvature is constant, given by
\begin{eqnarray}\label{}
{\cal{K}}=-\frac{g_{11}g_{22}-g_{12}^2}{g_{22}}.
\end{eqnarray}
 Now, assuming ${\cal{K}}<0$ and defining 
\begin{eqnarray}\label{k}
k\equiv-{\cal{K}},
\end{eqnarray}
the new coordinates $(\rho,\varphi)$ can be introduced by the  following coordinate redefinition 
\begin{eqnarray}\label{}
\begin{split}
x_1&=-L\,c_1\varphi-\sqrt{k}c_1\rho- \ln  \left( 2\,k\,{{\rm e}^{-{
				{2}{\sqrt {k}}}c_1 \rho}}+1 \right)  +\ln(k)+\frac{3}{2}\ln(2)
,
\\
x_2&={\frac {1}{{4 g_{22}}\,k}}\left(\sqrt {{2\,g_{22}}} \left( 2\,{{\rm e}^{-\sqrt {k}c_1\rho
		}}k-{{\rm e}^{\sqrt {k}c_1\rho}} \right) -4\,{g_{12}}\,
k\right)
,
\end{split}
\end{eqnarray}
where $L$ and $c_1$ are real  constants. Then, the line element \eqref{m1} recasts the following form
\begin{eqnarray}\label{m3}
ds^2\equiv g_{\mu\nu}^Rdx^\mu dx^\nu=-dt^2+c_1^2\left(d\rho^2+R^2(\rho)d\varphi^2\right),
\end{eqnarray}
with
\begin{eqnarray}\label{m4}
\begin{split}
R(\rho)&=\frac{\sqrt{2}L}{4\,k}\left(2\,k\,{{\rm e}^{-\sqrt{k}c_1{ \rho}}}+ {{\rm e}^{{
			{{ \sqrt{k} }c_1 \rho}}}}\right)=\frac{\sqrt{2}L}{2\,k}\cosh_{2k}(\sqrt{k} c_1 \rho).\\
\end{split}
\end{eqnarray}
The $\cosh_{2k}$ function is known as the deformed hyperbolic function introduced for the first time in \cite{ARAI199163,EGRIFES2000229}, in solving  Schrodinger equation with deformed potential.

In the case that the embedding of the space part of the above line element into ${R}^3$ is of interest, according to \eqref{canonicalpar},  the $\rho$ coordinate must  range in
\begin{equation} \label{rorange}
 \rho  \in  \left[{\frac {1}{\sqrt {k}c_1}\ln  \left( {\frac {\sqrt{2k}\,(a-1)}{Lc_1}} \right) },
 				\frac {1}{\sqrt {k}c_1}\ln  \left( {\frac {\sqrt{2k}\,(a+1)}{Lc_1}} \right) 
 \right],
\end{equation}
in which $a=\sqrt {{L}^{2}{{c_1}}^{2}+1}$, and hence
\begin{equation}\label{Rm}
R(\rho)\in \left[{\frac {L\sqrt {2} \left( 2\,k+1 \right) }{4k}},\frac{a}{\sqrt{k}c_1}\right].
\end{equation}
 These maximal circles, which appeared as a consequence of the Hilbert theorem \cite{eisenhart,14,15}, denote the non-removable singularities, i.e. the singular boundaries for the considered embedded surface, which are called the Hilbert horizons \cite{g}.
 
 The obtained line element \eqref{m3} with the $2$-dimensional space possessing Lie-symmetry can be considered as an especial class of metric \eqref{linelsurfrev} with $R(\rho)$ given by \eqref{solgaussian2}. It is worth noting that, although the Beltrami and elliptic type metrics can not be recovered from \eqref{m3}, the value of  $k=\frac{1}{2}$ can reduce the metric \eqref{m3} to a hyperbolic type case, described in \eqref{Rhyperbolic}. 

Now, motivated by the point that  the BTZ black hole solution has conformal relation to the spacetime with hyperbolic space geometry \cite{cvetivc2012graphene},   we are going to find a black hole solution, if exists, whose line element is conformally related to the metric \eqref{m3}. In this regard, we start with the metric $g_{\mu\nu}=h(\rho)g_{\mu\nu}^R$, i.e.
\begin{eqnarray}\label{m5}
ds^2=h(\rho)g_{\mu\nu}^Rdx^\mu dx^\nu
=h(\rho)\left(-dt^2+d\rho^2+R^2(\rho)d\varphi^2\right),
\end{eqnarray}
in which $R(\rho)$ is  $(2k)$-deformed hyperbolic function given by \eqref{m4} and $h(\rho)$ is an arbitrary function. 
The Gaussian curvature-related parameter $k$ \eqref{k} will be assumed to be an arbitrary positive constant.

To construct the black hole solutions, we  consider the $\beta$-function equations \eqref{tbetaGR}-\eqref{tbetafR}  with contributions of dilation, antisymmetric $B$-field, and central charge deficit $
\Lambda$,  which plays the role of a dilation potential in the string effective action \eqref{taction}. We start with finding the solutions to the  $\beta$-function equations at one loop order, i.e. $\alpha'=0$. Then, the solutions to the two-loop order of  $\beta$-function equations are considered.


	\subsection{$(2+1)$-dimensional black hole solutions of one-loop $\beta$-function equations  } \label{tsec3}
 First, we consider the $\beta$-function equations \eqref{tbetaGR}-\eqref{tbetafR} at the one-loop order, i.e. in the absence of the $\alpha'$ corrections.
 Considering \eqref{m5}, with the $R(\rho)$  given by \eqref{m4}, as the string frame metric   along with the  following form of field strength tensor of $B$-field
\begin{eqnarray}\label{tH}
H=\frac{1}{3!}E(\rho)\,dt\wedge d \rho\wedge d\varphi,
\end{eqnarray}
  the $(t,t)$, $(\rho,\rho)$, and $(\varphi,\varphi)$ components of   metric $\beta$-function  \eqref{tbetaGR}, $(t,\varphi)$ component  $B$-field $\beta$-functions  \eqref{tbetaBG}, and   dilaton $\beta$-functions \eqref{tbetafR}  
 yield following coupled set of differential equations  
 \begin{eqnarray}\label{ttt}
2{\ln h''}+{\ln h'}\, \left( {\ln (h\,R^2)'}+2\,\phi' \right) 
-2{ {{E}^{2}}{{h}^{-2}{R}^{-2}}}
 =0,
 \end{eqnarray}
  \begin{eqnarray}\label{trr}
2{
	\ln R''}+2{\ln h''}+2\phi''+{\ln R'}\,{\ln (h\,R^2)'}-{\ln h'}\,\phi'-{ {{E}^{2}}{{h}
		^{-2}{R}^{-2}}}
 =0,
 \end{eqnarray}
\begin{eqnarray}\label{t11}
2{\ln R''}+{\ln h''}+2{{\ln R'}}^{2}+2 \left( {\ln h'}+2\phi'
\right) {\ln R'}+\frac{1}{2}\,{{\ln h'}}^{2}+{\ln h'}\,\phi'-
\,{ {{E}^{2}}{{h}^{-2}{R}^{-2}}}
	=0,
\end{eqnarray}
\begin{eqnarray}\label{e(r)}
3\,{\ln h'}+2\,{\ln R'}-2\,\phi'-2\,{\ln E'}=0,
\end{eqnarray}
\begin{eqnarray}\label{tKt}
\begin{aligned}
4\,\phi''+4\,{\ln R''}+4\,{\ln h''}+2{\phi'}^{2}+2
{\ln (h\,R^2)'} \phi'+4\,{{\ln R'}}^{2}+2
\,{\ln h'}\,{\ln R'}+{{\ln h'}}^{2}-2\Lambda\,h-{ {{E
		}^{2}}{{h}^{-2}{R}^{-2}}}=0.
\end{aligned}
\end{eqnarray}
Here, the prime stands for derivative with respect to $\rho$.
Combining equations \eqref{ttt}-\eqref{tKt} one can obtain the following equation
\begin{eqnarray}\label{fi}
\begin{aligned}
2\phi''+2{\phi'}^{2}+ \left( 2\,{\ln R'}+{\ln h'} \right) \phi'-2\Lambda
h+{ {{E}^{2}}{{h}^{-2}{R}^{-2}}}
=0.
\end{aligned}
\end{eqnarray} 
Solving these equations, we obtain the following set of solution
\begin{eqnarray}\label{fi0}
\phi(\rho)={\phi_0},
\end{eqnarray}
\begin{eqnarray}\label{h}
h(\rho)=\frac{32k^2}{\Lambda}\left({{\rm e}^{{
				{ \sqrt{k} }c_1 \rho}}}-2\,k {{\rm e}^{{
				{{ -\sqrt{k}}c_1 \rho}}}}\right)^{-2},
\end{eqnarray}
\begin{eqnarray}\label{e00}
E(\rho)=\frac{32\,b\,k^2c_1^3}{\Lambda^{\frac{3}{2}}{\rm e}^{\phi_0}}\frac{\left(2k {{\rm e}^{{-
				{{ \sqrt{k} }c_1 \rho}}}}+{{\rm e}^{{{ \sqrt{k} }c_1 \rho}}}\right)}{\left(2k {{\rm e}^{{
				{{ -\sqrt{k} }c_1 \rho}}}}-{{\rm e}^{{
				{ \sqrt{k} }c_1 \rho}}}\right)^3},
\end{eqnarray}
where $\phi_{0 }$,  and $b$ are  real constant. Then,   \eqref{fi} lead the following constraint between the  constants of the solution
\begin{eqnarray}\label{con}
\Lambda\,L^2-b^2{\rm e}^{-2\phi_0}=0.
\end{eqnarray} 
Accordingly, the  metric \eqref{m5} recasts the following form as the solution of the one-loop $\beta$-function equations
\begin{eqnarray}\label{m6}
ds^2=\frac{32 c_1^2k^2}{\Lambda\left(2\,k\,{{\rm e}^{{{ -\sqrt{k} }c_1 \rho}}}- {{\rm e}^{{
				{{ \sqrt{k} }c_1 \rho}}}}\right)^2}\left(-dt^2+d\rho^2+\frac{b^2{\rm e}^{-2\phi_{0 }}}{32\Lambda k^2}\left(2\,k\,{{\rm e}^{{{ -\sqrt{k} }c_1 \rho}}}+{{\rm e}^{{
			{{ \sqrt{k} }c_1 \rho}}}}\right)^2d\varphi^2\right).
\end{eqnarray}
 In the special case where the Gaussian curvature equals $-\frac{1}{2}$, i.e. $k=\frac{1}{2}$, the metric \eqref{m6} can be compared to \eqref{BTZ} that represents the line element to which the BTZ solution is conformally related. In this case,  \eqref{m6}  reads 
\begin{eqnarray}\label{bt}
ds^2=\frac{2c_1^2}{\Lambda\sinh^2\left(\frac{\sqrt{2}}{2}c_1 \rho\right)}\left(-dt^2+d\rho^2+\frac{b^2{\rm e}^{-2\phi_{0 }}}{2\Lambda}\cosh^2\left(\frac{\sqrt{2}}{2}c_1 \rho\right)d\varphi^2\right).
\end{eqnarray}
  Although the dependence on the Gaussian curvature constant is absent in \eqref{bt}, the remaining constant $c_1$ can be considered to be related to the BTZ mass by $c_1=\sqrt{2 m}$. The conformal factor in \eqref{bt} diverges at $\rho=0$. However,  it can be excluded  from the $\rho$ coordinate range
in \eqref{Rhyperbolic} while embedding it into $R^3$,
considering  $\rho  \in  (- {\rm arccosh} (\sqrt{1 + \frac{1}{k\,c^2}}) ,0)$, which leads to $r>0$ in the BTZ metric \eqref{BTZ}.

Here, we continue with the  general solution \eqref{m6} to obtain the black hole metric, applying the following redefinition
\begin{eqnarray}\label{ror}
-\frac{\sqrt{32} c_1k}{\sqrt{\Lambda}\left(2\,k\,{{\rm e}^{{{ -\sqrt{k} }c_1 \rho}}}- {{\rm e}^{{
				{{\sqrt{k} }c_1 \rho}}}}\right)}d\rho=dr.
\end{eqnarray}
Since the metric \eqref{m6} has  diverging conformal factor at \begin{eqnarray}\label{ro0}
\rho_0=\frac{\ln(2k)}{\sqrt{k}c_1}, 
\end{eqnarray}
to have a well-defined coordinate redefinition, we can restrict the range of this coordinate to
 $\rho<\rho_0$ to have $r>0$.\footnote{If the embedding into $R^3$ is of interest, the $\rho_0$  belongs to the $\rho$-coordinate range of \eqref{rorange}.} 
Then, \eqref{ror} gives the $\rho$ coordinate in terms of new $r$ coordinate as follows
\begin{eqnarray}\label{ror1}
\rho={\frac {1}{2\sqrt{k} {c_1} }\ln  \left( {\frac {
			\left( r-c_2 \right) \Lambda+8\,\sqrt{k} {c_1} }{2 \left( r-c_2
			\right) \Lambda\,k}} \right) },
\end{eqnarray}
in which $c_2$ is an integrating constant. 

 Now, applying the conformal transformation \eqref{tcon} on \eqref{m6}  gives the not-yet finalized form of the Einstein-frame black hole metric in terms of $r$ as follows
\begin{eqnarray}\label{m7}
ds^2_{E}=\tilde{	g}_{\mu\nu}^Rdx^\mu dx^\nu= {{\rm e}^{2\,\phi_0}}(-F_0(r)
dt^2+F_0^{-1}(r)dr^2+W^2(r)d\varphi^2),
\end{eqnarray}
where defining $B(r)=\frac{\Lambda}{4} \left( r -{c_2} +8\,\sqrt{k}c_1  
\right)$, we have
\begin{eqnarray}\label{}
F_0(r)=B(r)\left( r-{c_2} \right),\label{f} \label{F00}\\ W^2(r)={{\rm e}^{-2\phi_0}}{\frac {b^{2} }{{\Lambda}^{2}k}}\left( B(r)-\sqrt{k}{c_1} \right) ^{2}\label{R}.
\end{eqnarray}
 Also,  the field strength tensor \eqref{tH} is 
 \begin{eqnarray}\label{H}
 H(r)=-\frac{1}{3!}\sqrt{\Lambda}\,W(r)\,dt\wedge dr\wedge d\varphi.
 \end{eqnarray}

 \section{Thermodynamic behavior of the $(2+1)$-dimensional black hole solutions}\label{thermo}
 Since the only physical quantity characterizing the obtained black hole solution is its mass,  one physical integrating constant needs to be present in the solution.
Noting \eqref{m7}-\eqref{H}, $b$  can be absorbed in a redefinition of the $\varphi$ coordinate.  To decide on the $c_1$ and $c_2$, 
we  investigate the thermodynamic quantities and satisfaction of the first law of black hole thermodynamics. 

 The solution is associated with the horizons located at $r_{h_1}=c_2-{\frac {8{\sqrt{k}} }{\Lambda}} c_1$ and $r_{h_2}=c_2$, where $r_{h_1}<r_{h_2}$.  
The Hawking temperature can be derived from the Euclidean regularity methods \cite{Hawking1983},   by\footnote{The same result can be obtained via the surface gravity definition   
	\begin{eqnarray}\label{}
	\kappa^2_h=-\frac{1}{2}{(\nabla_{\mu}{\xi}_{\nu})\left(\nabla^{\mu}{\xi}^{\nu}\right)}|_{r=r_{H}}=\tilde{	g}^{rr}\left(\sqrt{\tilde{	g}_{tt}}\right)'_{r=r_{H}}=k\,c_1^2,
	\end{eqnarray}
	where $\xi$  is the Killing vector normal to the horizon such that $T=\kappa_h/2\pi$.}
\begin{eqnarray}\label{T}
\begin{aligned}
T=&\frac{\sqrt{(\tilde{g}_{rr}^{-1}) '\tilde{g}_{tt}'}|_{r=r_{h_2}}}{4\pi },
\end{aligned}
\end{eqnarray}
which gives the temperature of the leading order solution \eqref{m7} by
\begin{eqnarray}\label{T00}
\begin{aligned}
T_0=\frac{ \sqrt{k} c_1}{2\pi}.
\end{aligned}
\end{eqnarray}
Furthermore, the black hole entropy can be obtained using the Wald’s formula \cite{wald}
\begin{eqnarray}\label{walds}
\begin{aligned}
S=-2\pi\int_{h} d\varphi \sqrt{\tilde{\sigma}} \frac{\delta \cal{L}}{\delta \tilde{	R}_{\mu\nu\rho\sigma}}\epsilon_{\mu\nu}\epsilon_{\rho\sigma},
\end{aligned}
\end{eqnarray}
where the $h$ denotes evaluating the integral on the horizon, $\tilde{\sigma}$ is the determinant of the metric of the $1$-dimensional boundary in the Einstein-frame, and $\epsilon_{\mu\nu}$ is the binormal to the horizon whose normalization is chosen as $\epsilon_{\mu\nu}\epsilon^{\mu\nu}=-2$. At the leading order of Einstein-frame string effective action \eqref{tGBaction}, i.e. $\alpha'=0$, we obtain
\begin{eqnarray}\label{S00}
\begin{aligned}
S_0 =\frac{2\pi\,b\,c_1}{\kappa_3^2\Lambda}.
\end{aligned}
\end{eqnarray}

 The solution  \eqref{m7} is not asymptotically flat and similar to BTZ black holes, whose asymptotic Ricci scalar is constant, the Einstein-frame Ricci scalar is a constant given by $\tilde{	R}=-\frac{3}{2}{\rm e}^{-2\phi_{0 }}\Lambda$. The  conserved mass of this non-asymptotically flat solution can be calculated via the quasi-local formalism  by
 \cite{PhysRevD471407}
 \begin{eqnarray}\label{tquasi}
m= \frac{1}{\kappa_3^2}\int_{B} d\varphi  \sqrt{\tilde{\sigma}}n^{a}\tau_{ab}\xi^{b},
 \end{eqnarray}
 in which $\kappa_3$ is the $3$-dimensional Newton constant, $B$ is the $1$-dimensional boundary, and $n^a$ is the time-like unit normal vector to the boundary  $B$. Also, noting that with the constant dilation \eqref{fi0}, the central charge deficit term  in the leading order of string effective action \eqref{tGBaction}  reduces to the negative cosmological constant term,  the  quasi-local stress tensor  is given by \cite{Dehghani}
 \begin{eqnarray}\label{tau0}
 {\tau_{ab}}=K_{ab}-\tilde{	h}_{ab}K-\frac{1}{l}\tilde{	h}_{ab},
 \end{eqnarray}
 where $K_{ab}$ is  the  extrinsic curvature of the $2$-dimensional boundary $\partial{\cal{M}}$ with induced metric $\tilde{	h}_{ab}$,  $K$ is the trace of $K_{ab}$, and on this $(2+1)$-dimensional spacetime 
  \begin{eqnarray}\label{l}
  l=\frac{2{\rm e}^{\phi_{0 }}}{\sqrt{\Lambda}}\end{eqnarray}

  The calculation of the mass is provided in the appendix A, where it is shown that \eqref{tquasi} leads to the mass expression  \eqref{mm}, which yields the mass of the zeroth order $\alpha'$ solutions by
 \begin{eqnarray}\label{m00}
 m_0={\frac {{b}\,c_2  }{32 \sqrt{k} }}\left( -c_2\Lambda+8\,\sqrt{k}{ c_1} 
 \right)+{\frac {{b}\,\sqrt {2} {{\rm e}^{{\phi_{0 }}}}}{32\Lambda\, \sqrt{k} }}\left( c_2\Lambda-4\,\sqrt {k
 } c_1\right) ^{2}
 .
 \end{eqnarray} 

 As mentioned before, 
  only one constant that characterizes the mass of the black hole is required to be present in the solution.  
Noting \eqref{T00}, \eqref{S00}, and \eqref{m00},
 if we consider  $c_2$ to be functionally related to $c_1$, i.e. $c_2(c_1)$,\footnote{{The functional relation between the appeared constant in the solutions has been for instance pointed out in  AdS context in \cite{Hertog}, where in the Einstein-Scalar models with the scalar filed of the form $\phi=\frac{\alpha}{r}+\frac{\beta}{r}$, the integrability of energy in Hamiltonian formalism, which contains $\delta Q_{\phi}=\int \beta \delta \alpha d\Omega+\dots$ term, requires a functional relation between the $\alpha$ and $\beta$ coefficients in the asymptotic expansion of the scalar field.}}  then
 the satisfaction of the first law of thermodynamics $dm_0=T_0\,dS_0$ fixes this relation as 
\begin{eqnarray}\label{c2}
c_2=\frac{4c_1\sqrt{k}}{\Lambda},
\end{eqnarray}
where
 the mass of these black hole solutions, recasts the following form
 \begin{eqnarray}\label{m0}
 m_0={\frac {{b} \sqrt{k}  }{2\Lambda
 }}c_1^2.
 \end{eqnarray}
 
Using \eqref{c2} and \eqref{m0} in \eqref{m7},  the final form of the black hole solutions  for one-loop $\beta$-function equation in the Einstein-frame reads  
 \begin{eqnarray}\label{mf}
 ds^2_{E}= {{\rm e}^{2\,\phi_0}}\left(-f_0(r)
 dt^2+f_0(r)^{-1}dr^2+{\frac {{b^2}\,{{\rm e}^{-2\phi_0}}  }{16\, {k}  }}r^2d\varphi^2\right),
 \end{eqnarray}
where
 \begin{eqnarray}\label{f0}
f_0(r)=\left(\frac{1}{4}\Lambda\,{r}^{2}-\frac{4kc_1^2}{\Lambda}\right)=\left(\frac{1}{4}\Lambda\,{r}^{2}-\frac{8\sqrt{k}}{b}m_0\right)
 \end{eqnarray}
where in the last expression the \eqref{m0} has been used. The obtained metric has constant curvature invariants without any essential singularity.
Also, the antisymmetric $B$-field associated with the filed strength tensor  \eqref{H}  becomes
\begin{eqnarray}\label{}
B_{t\varphi}=-{\frac {{b}\sqrt{\Lambda} }{2 \sqrt{k}  }}\,{{\rm e}^{-\phi_0}} r^2.
\end{eqnarray}

Compared to the BTZ solution to string effective action, obtained in \cite{Horowitz}, which was constructed with $\phi_0=0$ and including no integrating constant associated with the $B$-field $\beta$-function equation, the obtained solution \eqref{mf} contains the constants $\phi_0$ and $b$  along with the constant $k$,  related to the Gaussian curvature of the homogeneous space.
One may try to recover the BTZ black hole form by eliminating the $b$, $\phi_0$, and $k$  dependent factors the line element \eqref{mf} with  the coordinates redefinition
\begin{eqnarray}
r\rightarrow {\rm e}^{-\phi_0}\,r,\quad t\rightarrow {\rm e}^{\phi_0}\,r,\quad \varphi\rightarrow \frac{4k{\rm e}^{-\phi_0}}{b}\,\varphi,
\end{eqnarray}
where considering \eqref{l},  leads to
\begin{eqnarray}\label{mff}
ds^2_{E}= -\left(\frac{r^2}{l}-\frac{8\sqrt{k}}{b}m_0\right)
dt^2+\left(\frac{r^2}{l}-\frac{8\sqrt{k}}{b}m_0\right)^{-1}dr^2+r^2d\varphi^2,
\end{eqnarray}
noting that,  the mass defined by the expression  \eqref{tquasi}, being indifferent to the coordinate basis as a physical quantity,  is again given by \eqref{m0} for \eqref{mff}. Therefore, if one is interested in calling this metric as a BTZ solution, the price to pay will be the calling of the $\frac{8\sqrt{k}}{b}m_0$ factor in the metric as a mass, which is not the physical one. Hence, respecting the remaining $k$ and $b$-dependent factor in \eqref{mff} we consider the solution as a BTZ-like solution.
The BTZ black hole \eqref{BTZ} can be recovered from the solution \eqref{mff} by setting $b=8\sqrt{k}$.

In other words, we arrived at a BTZ-type solution starting from the spacetime conformal related to the line element whose spatial part is a homogeneous space with arbitrary negative Gaussian curvature ${\cal K}=-k$. The conformal relation of BTZ solution to  hyperbolic space has been addressed in \cite{g,cvetivc2012graphene}. Here, as mentioned before, the considered homogeneous spacetime is  of the deformed hyperbolic type and we showed that this more general Lobachevsky-type geometry with the arbitrary Gaussian curvature  has a conformal relation to the BTZ-like solution of string theory.

To see the thermal stability of the solutions, one can consider the specific heat capacity at constant pressure, given by 
\begin{eqnarray}\label{C}
C=T\left(\frac{\partial S}{\partial T}\right)={\frac {2\,\pi\,{b}\,{c_1}}{\Lambda}},
\end{eqnarray}
which is positive for $c_1>0$, indicating the thermodynamic stability of the solutions.

 It is worth putting the event horizons in contact with the notion of Hilbert horizon of the homogeneous spacetime described by metric \eqref{m3}. 
 The $c_2$ expression \eqref{c2} leaves only the $r_{h_2}$ as the positive radius of event horizon, where noting \eqref{m0}\footnote{The other radius reads 
 	$
 	r_{h_1}=-\frac{4}{\Lambda} \sqrt{k}{ {{ c_1} }{}}	.
 	$}
 \begin{eqnarray}\label{rh}
 r_{h_2}=\frac{4}{\Lambda} \sqrt{k}{ {{ c_1} }{}}
 .
 \end{eqnarray}
  As we have seen, the singular boundaries for the surface  appear while embedding the surface into ${R}^3$, where the $\rho$ coordinate is restricted to be in the range of \eqref{rorange}.
  Using the redefinition  \eqref{ror},\footnote{This redefinition gives the $r$ coordinate in terms of $\rho$ as
	$$r={\frac {{
				4}\,\sqrt {k} \,	c_1\left( 2\,k{{\rm e}^{-2\,\rho  \,
					\sqrt {k}{c_1}}}+1 \right) }{\Lambda\, \left( 2\,k{{\rm e}^{-2\, \rho \,
					\sqrt {k}\,{c_1}}}-1 \right) }}.
	$$}
the two radius of Hilbert horizons defined by the $\rho$-coordinate bounds in embedding into $R^3$, are   
\begin{eqnarray}\label{Hh}
r_{Hh_1}=-\frac{4}{\Lambda} \sqrt{k}{ {{ c_1} }{}}- {\frac {4\,{L}^{2}{{c_1}}^{3}\sqrt{k}}{ \left( \sqrt {{L}^{2}{{
					c_1}}^{2}+1}+1 \right) \Lambda}}
,\quad r_{Hh_2}=-\frac{4}{\Lambda} \sqrt{k}{ {{ c_1} }{}}+ {\frac {4\,{L}^{2}{{c_1}}^{3}\sqrt{k}}{ \left( \sqrt {{L}^{2}{{
					c_1}}^{2}+1}-1 \right) \Lambda}}
\end{eqnarray}
 where,  as fixed by \eqref{con}, $L=b{\rm e}^{\phi_0}\Lambda^{-\frac{1}{2}}$. The $r_{Hh_1}$ and $r_{Hh_2}$ are associated to the upper and lower  bounds of  $\rho$ in \eqref{ror}, respectively. 
 It should be also noted that the coordinate redefinition \eqref{rorange} added an extra limit on  the $\rho$ coordinate by restricting it to $\rho<\rho_0$. 
 Hence, if the embedding of the $2$-dimensional surface into $R^3$ is considered in the solutions,  the $\rho$ coordinate  must belong to the range $
\rho  \in  \left[{\frac {1}{\sqrt {k}c_1}\ln  \left( {\frac {\sqrt{2}\,(a-1)}{L\sqrt {k}c_1}} \right) },\rho_0
\right]
$, which is equivalent to the $r$-coordinate that starting at $r_{Hh_2}$  goes to infinity as $\rho$ approaches $\rho_0$.
The $r_{Hh_1}<0$,  not belonging to this range of $\rho$, is not of our interest. 
   Now,  assuming the $L$  to take the form of $L=(c_1)^{-1}\sqrt{p-1}$, or equivalently  $b={\frac {\sqrt {\Lambda({p}-1)} {}}{{{\rm e}^{-\phi0}}{c_1
}}}
$, in which $p$ is a constant, the radius of event horizon  \eqref{rh} and Hilbert horizon \eqref{Hh} are related generally by 
\begin{eqnarray}\label{hh33}
 r_{Hh_2}-r_{h_{2}}=\frac{4\,kc_1}{\Lambda}(\sqrt{p}-1).
\end{eqnarray}
It shows that, similar to the case when the space part of BTZ solution is embedded in $R^3$ with usual hyperbolic geometry \cite{cvetivc2012graphene}, the geometry ends before reaching the event horizon of the black hole.
Also, \eqref{hh33} shows that the two kinds of horizons can not coincide unless  $c_1=0$, for which the entropy of the black hole vanishes.

\section{$\alpha'$-corrected $(2+1)$-dimensional black hole solutions for two-loop $\beta$-function equations in Gauss-Bonnet RS}\label{22}
The string frame Ricci scalar of the obtained solution for the one-loop $\beta$-function equations is $R=-\frac{3\Lambda}{2}$. Hence, noting   \eqref{lamda}, the solutions are at the high curvature limit $R\alpha'\gtrsim 1$, where the  $\alpha'$ corrections to the string effective action become significant. 
Here, aiming at investigating only the consequences of including the $\alpha'$ corrections and neglecting the quantum loop corrections, the string coupling is assumed to be weak, i.e. $g_s={\rm e}^{-\phi_0}\ll1$.
Then, the two-loop $\beta$-function equations \eqref{tbetaGR}, \eqref{tbetaBG}, and   \eqref{tbetafR}, equivalent to the equations of motion of  Gauss-Bonnet $\alpha'$-corrected string effective action \eqref{tGBaction}, are considered to be solved to find the  $\alpha'$-corrections to the obtained black hole solution. In doing so, we start with a string frame metric of the form
\begin{eqnarray}\label{m10}
ds^2= -F(r)  
dt^2+{\frac {1}{ h(r)}}{{dr}}^{2}+W^2(r)d\varphi^2,
\end{eqnarray}
and the $H$ filed strength \eqref{tH}, where
 $W^2(r)={\frac {{b^2}\,{{\rm e}^{-2\phi_0}}  }{16\, {k}  }}r^2$. Then, considering the   two-loop   $\beta$-functions \eqref{tbetaGR}-\eqref{tbetafR} at Gauss-Bonnet RS, 
the $(t,t)$, $(\rho,\rho)$ and $(\varphi,\varphi)$ components of metric  $\beta$-function, $(t,\varphi)$ component of $B$-field $\beta$-function, and the dilaton $\beta$-function
reduce to the following coupled set of  equations 
\begin{eqnarray}\label{b11two}
\begin{aligned}
&{\ln F''}+{\ln F'}\, \left( {\ln (\sqrt{Fh}W)'}+\phi'
\right) +{\frac {{E}^{2}}{F{W}^{2}}}
-{\alpha'}\bigg[\frac {{E}^{2}h}{F{W}^{2}} \bigg( {\ln F''}+\frac{1}{4}{\ln F'}
\, \left( {\ln \left({F^{3}hW^{6}E^{-2}}\right)'}
\right)  \bigg) \\
&+\frac{1}{2}h \bigg( {{\ln F''}}^{2}+\ln F''\ln F'\ln (Fh)'+\frac{1}{4}{{\ln F'
	}}^{2} \left( {{\ln F'}}^{2}+\ln h'^2+2\ln F'\ln h'+4{{\ln W'}}^{2}
	\right)  \bigg) 
-{\frac {3h{E}^{4}}{2F^2{W}^{4}}}
\bigg]=0,
\end{aligned}
\end{eqnarray}

\begin{eqnarray}\label{brr2wo}
\begin{aligned}
&2\phi''+{\ln F''}+2\,{\ln W''}+{{\ln F'}}^{2}+  {\ln h'} \left({\ln (\sqrt{F}W)'}+\phi'\right)+\frac{1}{2}\ln F'^2+2\ln W'^2 +{\frac {E}{F{W}^{2}}}\\&
+\frac{\alpha'}{2}\bigg[\frac {{E}^{2}}{{W}^{2}} \big(3 \ln F''+6 \ln W''-2 \ln E''-\ln h''-\frac{5}{4} \ln h'^2- \ln (E^{4}W^{-6}F^{-3})' \ln h'+\frac{1}{4} \ln F'^2-3 \ln E'^2\\
&- 3\ln(W E^{-2})' \ln F'+\ln W'\ln(WE^6)'
\big)
+{h}\bigg(-\ln F''^2-\frac{1}{2}\ln F' \ln (Fh)' \ln F''-\frac{1}{8}(4 \ln W'^2+ \ln F'^2) \ln h'^2\\
&-2(2  \ln W'^2+  \ln h'  \ln W'+ \ln W'')  \ln W''-(2  \ln W'^3+\frac{1}{4} \ln F'^3) \ln h'-2  \ln W'^4-\frac{1}{8} \ln F'^4 \bigg)
-{\frac {3{E}^{4}h}{2{W}^{4}F^2}}
\bigg]=0,
\end{aligned}
\end{eqnarray}

\begin{eqnarray}\label{b33two}
\begin{aligned}
&2\,{\ln W''}+2\,{\ln W'}\, \left( {\ln (FW)'}+\phi'\right) +{\frac {{E}^{2}}{F{W}^{2}}}
-{\alpha'}{}\biggl[{\frac {{E}^{2}}{{W}^{2}} \left( 2\,{\ln W''}+{\ln W'
	}\,  {\ln \left({\sqrt{F^3h}W^3E^{-1}}\right)'}
	  \right) }
\\
&+\frac{2F^2}{h}  \left( (( \ln W'\ln(W^2h)'+\ln W'') \ln W''+\frac{1}{4} \ln W'^2 (\ln F'^2+4 \ln (Wh') \ln W'+\ln h'^2))\right)  -{\frac {3h{E}^{4}}{4{W}^{4}F^3}}
\biggr]=0,
\end{aligned}
\end{eqnarray}
	\begin{eqnarray}\label{twofi}
	\begin{aligned}
&{\phi''}+ \left( {\ln (\sqrt{hF}W)'}-\frac{4}{5}{\phi'} \right) {\phi'}+{\frac {\Lambda}{5h}}+
{\frac {{E}^{2}}{5{W}^{2}F}}
-
\frac{\alpha'}{10}\biggl[\frac{hE^2}{FW^2}\bigg( -\frac{1}{2}{\ln E'}\,  {\ln \left({E^3h^4}{W^{-4}}\right)'}\\
& +{\ln F'
}\, \ln (EW^2h^2) +\frac{9}{8}\,{{\ln F'}}^{2}+\frac{9}{2}\,{{\ln W'}}^{2}-{
	\ln E''}+3\,{\ln F''}+6\,{\ln W''}-\frac{1}{2}\ln h''\frac{1}{8}\ln h'(32\ln W'-5\ln h') \bigg) \\
&-
3h \bigg(\frac{1}{2}\ln F''^2+\frac{1}{2} \ln F' \ln (hF)' \ln F''+2 (2 \ln W'^2+ \ln h'\ln W'+\ln W'') \ln W''+\frac{1}{4} \ln (hF^2)' \ln F'^3\\
&+\frac{1}{8} (4 \ln W'^2+\ln h'^2) \ln F'^2+( \ln (h W2)')^2 \ln W'^2
 \bigg) -{\frac {5h{E}^{4}}{16W^4F^2}}
\biggr]=0,
	\end{aligned}
	\end{eqnarray}
\begin{eqnarray}\label{betaB2}
\begin{aligned}
&{\ln W'}-{\phi'}-{\ln E'}+\frac{1}{2}\ln(Fh^{-1})'
+\frac{\alpha'}{64}{{\rm e}^{-\phi}} \ln (hE^2W^{-2}F^{-1})'\biggl[2 \left( {\ln W'}\,{\ln F'}+2\,{{
	\ln W'}}^{2}+2\,{\ln W''} \right) F\\
&-{ { \left( 2{\ln F''}+\ln f'\ln (hF)' \right) {F}h}{{W}^{-2}}}+
	{( {{{\ln F'}}^{2}+\ln (FW^{-2})'\ln h'-4\,{{\ln W'}}^{2}+2{\ln F''}-4{\ln W''}}){{W}^{-2}}}  
\biggr]=0,
	\end{aligned}\end{eqnarray}
where the prime stands for derivation with respect to $r$.	

Solving the set of equations of  \eqref{b11two}-\eqref{betaB2} to obtain the $\alpha'$ corrections to the one-loop solution \eqref{mf}, by respecting the geometry of the spacetime, to which the solutions have had Weyl relation, we find the following solutions
\begin{eqnarray}\label{m111}
ds^2= -f_0(r) (1+\alpha'c_3) 
dt^2+{\frac {1+\alpha'c_4}{f_0(r)}}{{dr}}^{2}+{\frac {{b^2}\,{{\rm e}^{-2\phi_0}}  }{16\, {k}  }}r^2d\varphi^2.
\end{eqnarray}
	\begin{eqnarray}\label{pert}
		E(r)=-{\frac {\sqrt{\Lambda}{b}  }{4\, \sqrt{k}  }}{{\rm e}^{-\phi_0}}r,
	\end{eqnarray}
	in which $f_0(r)$   is given by \eqref{f0} and
	\begin{eqnarray}\label{}
	\begin{aligned}
c_3&={\frac {\Lambda}{\lambda(92
		\,\eta-17)}}
\left(-46\,\eta-26+\sqrt {2116\,{\eta}^{2}-10304\,\eta+3022}\right)
	 ,\\
	c_4&=\Lambda \eta,
	\end{aligned}
\end{eqnarray}
where we have defined $\lambda=\alpha'\Lambda$, which according to \eqref{lamda} equals $\frac{46}{3}$. Also, $\eta\sim 0.08991143357$.\footnote{$\eta$ is the solution for equation $58790944\,{\eta}^{4}-5628560\,{\eta}^{3}-349094\,{\eta}^{2}+62952\,
	\eta-2589
=0$.}

To calculate the  $\alpha'$-corrected mass in the quasi-local formalism using \eqref{tquasi}, the boundary terms related to the $\alpha'$ corrections in the string effective action need to be considered to have a well-defined variational principle. For the action \eqref{taction}, only the Gauss-Bonnet term is associated with a non-vanishing boundary term,\footnote{The ${\rm e}^{{6}\phi}\tilde{R}^{\alpha\beta\rho\sigma}H_{\alpha\beta\lambda}H_{\rho\sigma}^{~~\lambda}$ in the effective action \eqref{taction} needs a boundary term of type
	$$\int_{\partial {\cal M}} d^2x \,n_{\mu} {\rm e}^{{6}\phi} h^{\sigma\rho}\nabla_{\sigma}\delta g_{\nu\lambda}H^{\gamma\lambda}_{~~\rho}H_{\gamma}^{~\mu\nu},$$ 
	which vanishes
	because if $\delta g_{\nu\lambda}$ assumed to be constant on the boundary, then $h^{\sigma\nu}\nabla_\sigma \delta g_{\mu\nu}$ denoting the projected covariant derivative to the boundary with $h^{\mu\nu}$ must vanish.} which  leads to a correction term to the quasi-local stress tensor $\tau_{ab}$ \eqref{tau0}, given by  \cite{BRIHAYE2009204}
	\begin{eqnarray}\label{q}
	\tau_{ab}=K_{ab}-\tilde{	h}_{ab}K+\frac{1}{l}\tilde{	h}_{ab}+\frac{\alpha'{\rm e}^{{2 \phi}{}}}{2}\left(Q_{ab}-Q \,h_{ab}\right),
	\end{eqnarray}
 written for $(2+1)$-dimensional spacetime, where
	\begin{eqnarray}
	Q_{ab}=2KK_{ac}K_b^{~c}-2K_{ac}K^{cd}K_{db}+K_{ab}\left(K_{cd}K^{cd}-K^2\right)+2K\tilde{	R}_{ab}+\tilde{	R}K_{ab}-2K^{cd}\tilde{	R}_{cadb}-4\tilde{	R}_{ac}K_b^{~c}.
	\end{eqnarray}
 The Riemann tensor of the boundary metric $\tilde{	h}_{ab}$ has no non-zero components. Also, the extrinsic curvature $K_{ab}$ calculated for the metric \eqref{m10} in Einstein-frame, obtained via conformal transformation \eqref{tcon}, is given by 
	\begin{eqnarray}
	K_{ab}={ {{{\rm e}^{\phi_0}}h^{-\frac{1}{2}}}{}}\left[ \begin {array}{cc} -\frac{1}{2}F' &0\\ \noalign{\medskip}0&WW ' \end {array} \right],
		\end{eqnarray}
which leads to	$Q_{ab}=0$.\footnote{In fact, the boundary term for Gauss-Bonnet action is given by $$I_{ct}=-\frac{1}{8\pi G}\int_{\partial {\cal {M}}}d^{d-1}x\sqrt{\gamma}\frac{d-3}{2(d-4)} \left(\sqrt{\frac{\alpha'}{2}(d-3)(d-4)\Lambda{\rm e}^{\phi}+1}-1\right)L_{GB},
	$$
which actually vanishes when $d=3$.} Then, the similar procedure provided in appendix A can be used to derive the $\alpha'$-corrected mass. 
The temperature can be calculated via the definition \eqref{T}. Furthermore, the $\alpha'$-corrected black hole entropy can be calculated using the Wald’s formula  \eqref{walds}
\begin{eqnarray}\label{twaldf2}
\begin{aligned}
S
&
=-\frac{\pi}{2\kappa_3^2}\int_{H} dx\, \sqrt{\tilde{\sigma}}\epsilon_{\mu\nu}\epsilon_{\rho\sigma}\bigg(\left(1+\frac{1}{2}{\alpha'{\rm e}^{{2\phi}}}\tilde{	R}\right)\left(\tilde{	g}^{\mu\rho}g^{\nu\sigma}-\tilde{	g}^{\mu\sigma}\tilde{	g}^{\nu\rho}\right)-\frac{1}{4}\alpha'{\rm e}^{{6\phi}}H^{\mu\nu}_{~~\xi}H^{\nu\sigma\xi}\\
&~~~~~~~~~~~~~~~~~+\alpha'{\rm e}^{{2\phi}{}}\left(R^{\mu\nu\rho\sigma}-\tilde{	g}^{\mu\rho}\tilde{	R}^{\nu\sigma}-\tilde{	g}^{\nu\sigma}\tilde{	R}^{\mu\rho}+\tilde{	g}^{\mu\sigma}\tilde{	R}^{\nu\rho}+\tilde{	g}^{\nu\rho}\tilde{	R}^{\mu\sigma}\right)
\bigg)\\
&=\frac{\pi}{2\,\kappa_3^2}\int_{H} dx  \sqrt{\tilde{\sigma}} \left(4+\alpha'
E^2W^{-2}\right)\\
&=\frac{2\pi}{\kappa_3^2}A_h\left(1+\frac{\lambda}{4}\right).
\end{aligned}
\end{eqnarray}
The mass, entropy, and temperature are then given as follows
\begin{eqnarray}\label{twoth}
m=\sqrt{\frac{1+\alpha'c_3}{1+\alpha'c_4}}m_0, \quad T=\sqrt{\frac{1+\alpha'c_3}{1+\alpha'c_4}}T_0,\quad S=\left(1+\frac{\alpha'\Lambda}{4}\right)S_0,
\end{eqnarray}
in which $m_0$, $T_0$, and $S_0$ are given by \eqref{m0}, \eqref{T00}, and \eqref{S00}. When $\alpha'\rightarrow 0$, the thermodynamic quantities of $\alpha'$-corrected solutions  \eqref{twoth} reduce to those of the leading order solution.

\section{Conclusion}\label{conclussion}
 We have constructed $(2+1)$-dimensional BTZ-like black hole solutions for string effective actions at leading order and the first order of $\alpha'$, involving the contributions of dilaton, anti-symmetric $B$-field, and central charge deficit $\Lambda$ that appeared in the role of a negative cosmological constant. 
 The basic idea of this work was 
considering the $(2+1)$-dimensional spacetime whose two-dimensional space part has the symmetries of two-dimensional Lie algebra, admitting a homogeneous metric.  The homogeneous surface has been shown to be of constant negative Gaussian curvature type surfaces, described by deformed hyperbolic function.  Investigating the embedding of the homogeneous surface on $R^3$, we found the boundaries of the spacetime. 
      
     Inspired by the point that the line element of BTZ black hole is Weyl-related to the line element of a spacetime 
   with hyperbolic space geometry \cite{cvetivc2012graphene}, we found a black hole solution for leading order of string effective action on the $(2+1)$-dimensional spacetime that is Weyl-related to the considered homogeneous spacetime.     
   Calculating the Wald entropy, temperature, and quasi-local mass,  the thermodynamic properties of the solutions have been investigated.    
   Generally, the leading order solutions and their thermodynamic quantities are given in terms of the constants, including  the central charge deficit $\Lambda$,  the  Gaussian scalar curvature dependent parameter $k$, constant dilaton $\phi_{0 }$,  the $B$-field $\beta$-function equation's integrating constant $b$,  and the two integrating constant $c_1$ and $c_2$. 
   According to the physical discussion presented in \cite{HAJIAN2017228}, $\phi_{0 }$ is not a physical parameter. The $c_2$, on the other hand, has been found to be functionally related to $c_1$  to satisfy the first law of thermodynamics. 
   Besides the mass and $\Lambda$, the obtained line element at the leading order of string effective action contains extra constants including the $\phi_0$, $k$, and $b$, where the latter two constants can not  be generally eliminated by redefinition of the coordinates. Having a similar structure to the BTZ solution but including the non removable extra $k$ and $b$ dependent factor,  the metric is refereed as a BTZ-like solution.  It can recast the form of the BTZ black hole by choosing a particular set of constants.

      The presence of the central charge deficit $\Lambda$, which is of order $\alpha'^{-1}$,  resulted in the high curvature of the leading order solutions with $R\alpha'\gtrsim1$. This characteristic underlines the requirement of considering the $\alpha'$ corrections in the effective action or equivalently the higher order   $\beta$-function equations.
     In these cases, the whole $\alpha'$ correction series is prescribed by the conformal invariance condition to be taken into account.
     Nevertheless, aimed at providing a pattern that may be obtained in the presence of $\alpha'$ corrections, we limited our calculation to the first order $\alpha'$ (two-loop) $\beta$-function equations as the equations of motion of the  $\alpha'$-corrected Gauss-Bonnet string effective action
     , which is valid at the weak string coupling limit. The mass, entropy, and temperature calculated for the $\alpha'$-corrected solutions showed that the first order $\alpha'$-corrections rescaled these thermodynamic quantities compared to those of the leading order solutions.
    
    Similar to the BTZ black hole, the solutions at both orders $\alpha'$ are regular with no curvature singularities. Furthermore, deriving the heat capacity, it has been shown that the solutions  have thermodynamic stability. 
Also, considering the singular boundaries of the negative Gaussian curvature pseudosphere, interpreted as the Hilbert horizons,  in relation to the event horizons of the obtained black hole it turned out that, similar to the  case that BTZ black hole are studied in relation to the hyperbolic geometry \cite{cvetivc2012graphene}, the embedded space part of the homogeneous spacetime into $R^3$   may end before reaching the event horizons. It has been seen that these two kinds of horizons can not generally coincide, unless when the mass of the black hole vanishes.
Also, it should be stressed that the metric is different from those considered in the so-called deformed hyperbolic black holes \cite{Deformed},  whose two-dimensional metric curvature invariants are not constant.

There are several venues worth exploring regarding the solutions presented here. The family of surfaces of constant negative Gaussian curvature has been one of the most fruitful classes of surfaces considered in the approaches that address the interrelation between various branches of physics. Particularly, they have been used in investigating the measurable effects of  QFT in curved spacetime describing curved graphene, as a  $2$-dimensional real system \cite{g,cvetivc2012graphene,Iorio:2012xg,Chen:2012uc,hasanirokh}.
Their results concern the cases of Beltrami, elliptic, and hyperbolic pseudosphere. In this context, the new solutions provided in this work can be considered for the sake of studying the measurable effects,
like the Hawking-Unruh effect, while the QFT and condensed matter concepts are taken into account.

\section*{Acknowledgment}
This research was supported by Azarbaijan Shahid Madani
University under Grant No. $ 1402/231-2$.

 \section*{Appendix A}\label{appa}
 In this appendix, we present the 
 calculation for  quasi-local mass. 
  With Killing vector $\bar{\xi}^{\mu}=\delta^{\mu}_{t}$,  \eqref{tquasi} leads to the  following mass expression for the leading order solution \eqref{m7}
 \begin{eqnarray}\label{tquasiii}
 m_0= \frac{1}{\kappa^2}\int_{B} d\varphi \sqrt{\tilde{\sigma}}\sqrt{F_0(r)}\,\varepsilon,
 \end{eqnarray}
 where, similar to the quasi-local mass of BTZ black holes calculated in 
 \cite{CHAN1996199,Brown}, $\varepsilon=K-\varepsilon_0$ in which $\varepsilon_0$ denoting the  zero of the energy, is a function of  background metric  and 
 \begin{eqnarray}\label{}
 K=-{W'(r)\sqrt{F_0(r)}}+\frac{{\rm e}^{\phi_{0 }}}{2}\sqrt{2\Lambda}W(r).
 \end{eqnarray}
 Then, the conserved mass is provided by
 \begin{eqnarray}\label{mm}
 m_0=\frac{b}{8 \sqrt{k} }\sqrt{F_0(r)}\,\left(\sqrt{F_{00}(r)}-\sqrt{F_0(r)}+\sqrt{\Lambda} {{\rm e}^{2\phi_0}}(W_{00}(r)-W(r))\right),
 \end{eqnarray}
 in which $F_0(r)$ and $W(r)$ are  given by \eqref{f} and \eqref{R}, respectively, and  $F_{00}(r)$ and $W_{0}(r)$ are the background metric functions, given by
 \begin{eqnarray}\label{}
 {F_{00}(r)}  =\frac{1}{4} {{r} 
 	\left( \Lambda\,{r}-2\,c_2\Lambda+8\, \sqrt{k}c_1  \right) },\quad W_{0}(r)=\frac{{{\rm e}^{-\phi_0}}}{4}{\frac {
 		{b}}{\sqrt {k\,\Lambda}}} ({r\,(-2\,\Lambda\,c_2+\Lambda\,r+8\,\sqrt {k}{c_1})}).
 \end{eqnarray}
 Accordingly, \eqref{mm} yields the mass
 \begin{eqnarray}\label{m000}
 m_0={\frac {{b}\,c_2  }{32 \sqrt{k} }}\left( -c_2\Lambda+8\,\sqrt{k}{ c_1} 
 \right)-{\frac {{b}\,\sqrt {2} {{\rm e}^{{\phi_{0 }}}}}{32\Lambda\, \sqrt{k} }}\left( c_2\Lambda-4\,\sqrt {k
 } c_1\right) ^{2}
 .
 \end{eqnarray}

	
\bibliographystyle{h-physrev}
\bibliography{blackhole22}
\end{document}